\documentclass[graybox]{svmult}

\usepackage{mathptmx}
\usepackage{helvet}
\usepackage{courier}
\usepackage{type1cm}

\usepackage{makeidx}
\usepackage{graphicx}

\usepackage{multicol}
\usepackage[bottom]{footmisc}

\usepackage{amssymb}
\usepackage{bm}
\usepackage{epsf}
\usepackage{graphicx}
\usepackage{cite}
\input colordvi

\newcommand{\be}{\begin{equation}}
\newcommand{\ee}{\end{equation}}
\newcommand{\bea}{\begin{eqnarray}}
\newcommand{\eea}{\end{eqnarray}}
\newcommand\ptl{\partial}

\makeindex

\begin{document}

\title*{ Plasma Perturbations and Cosmic Microwave Background Anisotropy in the
Linearly Expanding Milne-like  Universe}
\titlerunning{Plasma Perturbations and CMB Anisotropy in the
Milne-like  Universe}
\author{S.L. Cherkas and  V.L. Kalashnikov}
\authorrunning{Plasma Perturbations and CMB Anisotropy in the
Milne-like  Universe}

\institute{S.L. Cherkas \at Institute for Nuclear Problems,
Bobruiskaya str. 11, Minsk, 220050, Belarus,
\email{cherkas@inp.bsu.by} \and V.L. Kalashnikov \at Institute of
Photonics, Vienna University of Technology, Vienna A-1040, Austria
\email{vladimir.kalashnikov@tuwien.ac.at}}

\maketitle

\abstract*{We expose the scenarios of primordial baryon-photon plasma evolution within the framework of the Milne-like universe models. Recently, such models find a second wind and promise an inflation-free solution of a lot of cosmological puzzles including the cosmological constant one. Metric tensor perturbations are considered using the five-vectors theory of gravity admitting the Friedmann equation satisfied up to some constant. The Cosmic Microwave Background (CMB) spectrum is calculated qualitatively.}

\abstract{We expose the scenarios of primordial baryon-photon plasma evolution within the framework of the Milne-like universe models. Recently, such models find a second wind and promise an inflation-free solution of a lot of cosmological puzzles including the cosmological constant one. Metric tensor perturbations are considered using the five-vectors theory of gravity admitting the Friedmann equation satisfied up to some constant. The Cosmic Microwave Background (CMB) spectrum is calculated qualitatively.}

\section{Introduction}
\label{sec:1}

Present universe is transparent for photons, but it was not the same before the hydrogen recombination at the red-shifts of $z\approx 1100$ \footnote{$z$ is the red-shift parameter used as a measure of
cosmological time and distance: $z+1 = a_0/a(\eta)$, where $a_0$
is the present scale factor value, and $a(\eta)$ is the scale
factor at some earlier photon emission time $\tau$
\cite{mukh,CMB}.}, when it was filled with the photon-baryon plasma. Protons and electrons were coupled to the radiation through the Compton scattering by electrons which in turn are coupled to the baryons by Coulomb interaction \cite{mukh,dod}. Such primordial plasma perturbations were
 widely considered in cosmology, and their fingerprints depend on a law of the universe expansion that is the crucial point
 for our further analysis.

 Recently,
the Milne-like cosmologies considering the linearly expanding (in
cosmic time) universe models \cite{milne1,milne2} again attract an
attention \cite{loh1,loh2,levy,fm,fm2,sh,ben,isa,moncy,ling}.
Instead of the original open and empty Milne universe model
\cite{milne1,milne2} \footnote{The universe proposed initially by Milne describes an open and empty (i.e., Minkowski) spacetime which expands linearly
with time \cite{mukh,milne1,milne2}. It is negatively-curved spatially (i.e., hyperbolical in 3-dimensions) but is ``flat'' in 4 (i.e., spacetime)-dimensions.},
the flat universes filled with some exotic
matter are considered. It seems reasonable to associate such ``a primordial matter fluid'' with the vacuum \cite{peter}.

We will consider the perturbations of plasma consisting of
photons, baryons, and electrons in a linearly expanding
(Milne-like) universe with taking into account the metric tensor
and vacuum perturbations. Here, we will use the oversimplified
model of plasma as a \emph{pure radiation}, i.e., a substance with
the equation of state $w=1/3$ \footnote{We use a classical
definition for the equation of state parameter $w$ corresponding
to a perfect fluid, that is the ratio of pressure to density
\cite{melia}.}, to obtain an analytical solution. This
approximation is admissible because initially, the temperature is
sufficiently large to consider all the particles as a relativistic
fluid. Then, the particles decay eventually to the photons,
electrons, and baryons. According to observations number of
photons is of $10^9$ times larger than that of nucleons and
electrons. Thus, the nucleons contribute at only the late stage of
the universe evolution. We will base our analysis of the metric
tensor perturbations, which contribute to the primordial plasma
formation, on the five-vectors theory of gravity \cite{ann}. The
quantization of this model could resolve the problem of huge
vacuum energy \cite{w} and allow omitting its main part
\footnote{Below, the system of units $\hbar=c=1$ will be used, and
we define the present scale factor as $a_0=1$.}.

\section{Perturbations of Plasma and Vacuum}
\label{sec:2}

We expose the perturbation theory for primordial photon-baryon plasma, vacuum and metric tensor.
Vacuum issues the well-known challenge for quantum or, at least, semiclassical theory \cite{peter,w,birrel,z,adler}.
Here, we will consider a vacuum purely classically, that is as a substance producing the linear
expansion of the universe in the framework of the developed theory \cite{ann} which admits adding
or extracting some constant to the energy density.

\subsection{Underlying Gravity Theory}
\label{subsec:11}
The conventional theory of the CMB spectrum is the General
Relativity theory (GR) (e.g., see \cite{CMB}). In the case of the
Milne-like cosmology, the issue is more complicated, because an
origin of linear universe expansion is not clear. As was shown,
such linear expansion could arise from the residual vacuum
fluctuations of quantized fields including the scalar and
gravitational ones after omitting the main part of huge vacuum
energy \cite{peter}. As was mentioned above, the mystery of
cosmological vacuum is among the critical issues of modern physics
\cite{w,z,adler}. Below we will use the theory which validates
omitting the vacuum extra-energy and, besides, provides obtaining
the analytical solutions.

Let's start from the Einstein-Hilbert action for GR in the form of \cite{lan}:

\bea
S=-\frac{M_p^2}{12}\int \mathcal G\sqrt{-g}\,d^4x,
\label{sc}
\eea
where $\mathcal G=g^{\alpha \beta } \left(\Gamma _{\alpha \nu
}^{\rho }
   \Gamma _{\beta \rho }^{\nu }-\Gamma _{\alpha \beta
   }^{\nu } \Gamma _{\nu \rho }^{\rho }\right)$, and $M_p$ is the Planck mass,
    which is chosen as $M_p=\sqrt{\frac{3}{4\pi G}}$.

The next step is a violation of the general coordinate covariance
principle in (\ref{sc}) according to the Milne's perception of the
principally different concept of time in GR and quantum mechanics
\cite{kiefer,anderson,zeh}, so that we will consider the
restricted class of metrics $g_{\mu\nu}$ in the form of\be
ds^2\equiv g_{\mu\nu}dx^\mu dx^\nu =a^2\left(1-\ptl_m
P^m\right)^2d\eta^2-\gamma_{ij}(dx^i+N^id\eta)(dx^j+N^jd\eta),
\label{int}
\ee
where $\gamma_{ij}$ is the induced three metric, $a=\gamma^{1/6}$
is the scale factor defined locally, and $\gamma=\det \gamma_{ij}$. A
spatial part of the interval (\ref{int}) can be written as
\be
dl^2\equiv\gamma_{ij}dx^idx^j=a^2(\eta,\bm x)\tilde
\gamma_{ij}dx^idx^j,
\ee
where $\tilde \gamma_{ij}=\gamma_{ij}/a^2$ is a matrix with the
unit determinant.
  The
interval (\ref{int}) is analogous to the the ADM one \cite{adm},
but the expression $1-\ptl_m P^m$ is used instead of a lapse
function, where $\ptl_m$ is a partial derivative and $P^m$ is a three-vector. Varying the action over vectors $\bm P$, $\bm N$ and three
metric $\gamma_{ij}$ leads to the equations of the five-vectors theory (FVT) \cite{ann}:

\bea
\frac{\ptl g^{\mu\nu}}{\ptl\gamma_{ij}}\left(\frac{(\ptl \mathcal
G\sqrt{-g})}{\ptl g^{\mu\nu}}-\frac{\ptl}{\ptl
x^\lambda}\frac{\ptl (\mathcal G\sqrt{-g})}{\ptl ( \ptl_\lambda
g^{\mu\nu})}
-\frac{6}{M_p^2}T_{\mu\nu}\sqrt{-g}\right)=0,\nonumber\\
\frac{\ptl g^{\mu\nu}}{\ptl N^{i}}\left(\frac{\ptl (\mathcal
G\sqrt{-g})}{\ptl g^{\mu\nu}}-\frac{\ptl}{\ptl
x^\lambda}\frac{\ptl (\mathcal G\sqrt{-g})}{\ptl ( \ptl_\lambda
g^{\mu\nu})}
-\frac{6}{M_p^2}T_{\mu\nu}\sqrt{-g}\right)=0,\nonumber\\
\frac{\ptl g^{\mu\nu}}{\ptl( \ptl_j P^{i})}\frac{\ptl}{\ptl
x^j}\left(\frac{\ptl (\mathcal G\sqrt{-g})}{\ptl
g^{\mu\nu}}-\frac{\ptl}{\ptl x^\lambda}\frac{\ptl( \mathcal
G\sqrt{-g})}{\ptl ( \ptl_\lambda g^{\mu\nu})}
-\frac{6}{M_p^2}T_{\mu\nu}\sqrt{-g}\right)=0.\label{eqgr}
\eea
Eqs. (\ref{eqgr}) are \emph{weaker} than the GR ones.  At the same time,  the restrictions $\bm \nabla
(\bm \nabla\cdot \bm P)=0$ and $\bm \nabla (\bm \nabla\cdot \bm
N)=0$ on the Lagrange multipliers arise \cite{ann}.
In the particular case of $\bm \nabla \cdot \bm N=0$, \emph{the Hamiltonian constraint is satisfied up to some constant}.

The next step is to develop a theory for the scalar perturbations in
the gauge of $\bm P=0$, $\bm N=0$:
\be
ds^2=a(\eta)^2(1+2A)\left(d\eta^2-\left(\left(1+\frac{1}{3}\sum_{m=1}^3
\ptl_m^2F\right)\delta_{ij}-\ptl_i\ptl_jF\right)dx^idx^j\right).
\label{int1}
\ee
An interval (\ref{int1}) is a particular form of the interval (\ref{int}) up to the higher order terms in $F(\eta,\bm
x)$ by virtue of
\[\ln\left[\det\left(\left(1+\frac{1}{3}\sum_m
\ptl_m^2F\right)\delta_{ij}-\ptl_i\ptl_jF\right)\right]\approx \mbox{tr}
\left(\left(\frac{1}{3}\sum_m
\ptl_m^2F\right)\delta_{ij}-\ptl_i\ptl_jF\right)=0.
\]
Writing Eqs. (\ref{eqgr}) up to the first order relatively $A(\bm x,\eta)$ and $F(\bm x,\eta)$ leads
to the required perturbation theory.

\subsection{Energy-Momentum Tensor}
\label{subsec:12}

As was above mentioned, we violate eventually the general
coordinates' transformation invariance by the restriction of the
metrics' class by representing them in the form of (\ref{int}). To
built the energy-momentum tensor in the field theory, one should
write the corresponding special relativistic expression and then
change the partial derivatives to covariant ones. Using a
hydrodynamic approximation is more convenient. In this framework
the energy-momentum tensor is
\be
T_{\mu\nu}=(p+\rho)u_{\mu}u_{\nu}-p\, g_{\mu\nu}.
\label{tmn}
\ee
The equations of motion for some fluid in the GR can be obtained from both the equations
of motion of the fluid point-like components and the conservation of the energy-momentum
tensor $D_\mu T^{\mu\nu}=0$, where $D_\mu$ is a covariant derivative. In FVT, the energy-momentum
tensor conserves only in the Minkowski space-time. However, one can deduce the equation of motion for
fluid from the conservation of energy-momentum tensor by virtue of the Eqs. (\ref{eqgr}) self-consistency
in the particular gauge (\ref{int1}). Below, we will consider the scalar perturbations of a fluid $c$ (the index $c$ denotes a kind
of fluid) in the
form of $\rho_c(\eta,\bm x)=\rho_c(\eta)+\delta \rho_c(\eta,\bm
x)$, $p_c(\eta,\bm x)=p_c(\eta)+\delta p_c(\eta,\bm x)$ and
represent the 4-velocity in the form of
\be
u^{\mu}_c=\frac{1}{a(\eta)}\{(1-A),\bm \nabla v_c(\eta,\bm x)\},
\label{vr}
\ee
where $v_c(\eta,\bm x)$ is a scalar function.

\subsection{Zero-Order Equations}
\label{subsec:13}

  The zero-order evolution equation for logarithm of the scale factor $\alpha(\eta)=\ln a(\eta)$
 takes the form
\be
\alpha ''+ \alpha^{\prime 2}=M_p^{-2} e^{2 \alpha} (\rho -3 p),
\label{em0}
\ee
where  $\rho=\sum_c \rho_c$ and $p=\sum_c p_c$ are uniform
energy density and pressure, respectively. Summation is performed
over all the kinds of matter, but here we will consider only
vacuum $c=v$ and radiation $c=r$. For every component of a substance, the equation of motion is:
\be
\rho_c '+3 \alpha ' (\rho_c +p_c)=0.
\label{em1}
\ee
Pressure of a fluid is connected with the energy density as $p_c=w_c\rho_c$ (see the footnote 3 above and Ref. \cite{melia}).
It is worth mentioning that the Friedmann equation is satisfied only up to some constant in the framework of the model considered:

\be
M_p^{-2} e^{4 \alpha } \rho (\eta )-\frac{1}{2} e^{2 \alpha}
\alpha ^{\prime 2}=const,
\label{Fried}
\ee
that is the integral of motion of Eqs. (\ref{em0}), (\ref{em1}).

As was shown \cite{our1}, the residual vacuum fluctuations can
explain a nearly-linear universe expansion. Here, for simplicity,
we will use an empirical consideration. Let us analyze a linear
universe expansion that means $a(\eta)=B\exp\left(\mathcal H\,
\eta\right)$ in conformal time, and find the corresponding
empirical equation for the vacuum state. The very simple equation
of state arises if we set a constant in the Friedmann equation
(\ref{Fried}) so that
\be
M_p^{-2} e^{4 \alpha } \rho_v -\frac{1}{2} e^{2 \alpha} \alpha
^{\prime 2}=0.
\label{fr0}
\ee
It is possible because $\rho_r e^{4 \alpha }$ is also constant.
Under such choice of a constant, the equation of the vacuum state
will be $w_v=-1/3$. This equation of state is widely discussed
earlier \cite{fm,fm2,moncy}. One may obtain from Eq. (\ref{em1}) $\rho_v e^{2 \alpha }=const$ for
the vacuum, that results in (see Eq. (\ref{fr0})):
\be
a(\eta)=\exp\left( \alpha(\eta)\right)=B\exp\left(\mathcal H
\eta\right),
\label{aeta}
\ee
where $B$ is some constant. In the cosmic time $dt=a(\eta)d \eta$
\be
a(t)=\mathcal H t,
\ee
i.e., it is a linear expansion of the universe.

\subsection{Perturbations}
\label{subsec:14}

  Introducing the quantity $V_{c}=(p_c+\rho_c)v_{c}$ for every fluid $c$ and expanding all perturbations into the Fourier series $\delta \rho_c(\bm
x)=\sum_{\bm k}\delta\rho_{c\bm k}e^{i \bm k \bm x }...$ etc. result in the equations for perturbations:

\bea
-6 A_{\bm k}'+6 A_{\bm k} \alpha '+k^2 F_{\bm k}'+\frac{18}{M_p^2}
e^{2 \alpha } \sum_cV_{c \bm k}
=0,\label{con1}\\
 -18 \alpha ' A_{\bm k}'-18 A_{\bm k} \alpha ^{\prime 2}-6 k^2A_{\bm k}
+k^4F_{\bm k} +\frac{18}{M_p^2} e^{2 \alpha } \sum_c \delta
\rho_{c\bm k}+4 A_{\bm k}\, \rho_c
=0,\label{con2}\\
-12 A_{\bm k}-3 \left(F_{\bm k}''+2 \alpha ' F_{\bm k}'\right)+k^2
F_{\bm k}=0,\\
-9 \left(A_{\bm k}''+2 \alpha' A_{\bm k}'\right)-18 A_{\bm k}
\alpha''-18 A_{\bm k} \alpha^{\prime 2}-9 k^2 A_{\bm k} +k^4
F_{\bm k} \nonumber\\-\frac{9}{M_p^2} e^{2 \alpha }\sum_c 4 A_{\bm
k} (3 p_c-\rho_c)+3 \delta p_{c\bm k}-\delta \rho_{c\bm k}=0,\\-3
\alpha' ( \delta p_{c\bm k}+\delta \rho_{c\bm k})-3 A_{\bm k}'
(\rho_{c}+p_{c})-\delta \rho_{c\bm k}'+k^2 V_{c\bm k}=0,\\
(\rho_{c}+p_{c}) A_{\bm k}+4 V_{c\bm k} \alpha '+\delta p_{c\bm
k}+V_{c\bm k}'=0.\label{lasteq}
\eea
The last two equations, obtained from the energy-momentum
conservation, are assumed to be valid for every $c$-substance
under consideration. The choice of the constant in Eq.
(\ref{Fried}) is arbitrary. The constraint equations (\ref{con1})
and (\ref{con2}) are consistent with the other equation under this
arbitrary choice.It is not true in a perturbation theory within the framework of GR, where a perturbation of the constraint equations is consistent with other equations only if a sum of the mean densities of all fluids equals the critical density (for the flat universe). Here we consider the flat universe in a mean, but the sum of the mean densities is determined up to some constant, and nevertheless, all the equations for perturbations are self-consistent. With that chosen constant in Eq. (\ref{Fried}), the radiation does not affect
the universe expansion and the equation of state $w_v=-1/3$ for the vacuum results in linear expansion of the universe.
 Thus, the equations of state are
 $w_v=-1/3$ for the vacuum and $w_r=1/3$ for the
radiation \footnote{$\delta p_{c\bm k}=w_c\delta
\rho_{c\bm k}$ is assumed, as well.}.

Such choice of the constant in (\ref{Fried}) is an invention
expired by the existence of the analytical solution in this case.
The above system of the equations can be reduced to a single
linear equation with the constant coefficients under the assumption
of $a(\eta)=B\exp\left (\mathcal H \eta\right)$ and
$\rho_r=\frac{\rho_{r0}}{a^4(\eta)}$, where $\rho_{r0}$ is a density
of radiation at the present time:
\bea
9 \delta \rho_{r\bm k}^{(4)}+6 \left(30\mathcal H \delta
\rho_{r\bm k}^{(3)}+\left(222 \mathcal H^2+k^2\right) \delta
\rho_{r\bm k}''+10 \mathcal H \left(72 \mathcal H^2+k^2\right)
\delta \rho_{r\bm k}'\right)\nonumber\\+\left(48 \mathcal
H^2+k^2\right) \left(108 \mathcal H^2+k^2\right)\delta \rho _{r\bm
k} =0.
\eea

\noindent That allows obtaining the solution for the perturbation of radiation density:
\bea
\delta \rho_{r\bm k}=e^{-6 \eta  \mathcal H}\left(C_1e^{-i \frac{
\eta k}{\sqrt{3}}}+C_2e^{i \frac{\eta k}{\sqrt{3}}}\right)+e^{-4
\eta \mathcal H}\left(C_3e^{-i \frac{ \eta k}{\sqrt{3}}}+C_4e^{i
\frac{\eta k}{\sqrt{3}}}\right).
\eea

\noindent For a ``flux'' of the radiation fluid $V_{r\bm k}$, we
have
\bea
V_{r\bm k}=\frac{B^2 \mathcal H {M_p}^2} {6 k {\rho_{r0}}}e^{-4
\eta \mathcal H} \left({C_1} \left(k-i \sqrt{3} \mathcal
H\right)e^{-i \frac{ \eta k}{\sqrt{3}}}+{C_2} \left(k+i \sqrt{3}
\mathcal H\right) e^{\frac{i \eta k}{\sqrt{3}}}\right).
\eea

\noindent Other functions $A_{\bm k}$, $F_{\bm k}$, $\delta \rho_{v\bm
k}$, $V_{v\bm k}$ found from the system (\ref{con1})-
(\ref{lasteq}) are presented in Appendix.

The constants $C_1,C_2,C_3,C_4$ have to be determined from the
initial conditions. The constants $Z_1, Z_2$ (see Appendix) do not
contribute to the radiation density perturbations. Thus, we will
equal them to zero. Indeed, it is reasonable to assume that an
empty universe (i.e., filled by the only vacuum) has no any rising
physical perturbation, and only perturbations connected with the
radiation over the vacuum have a physical meaning. For simplicity,
we assume that the only perturbations
 of radiation density $\delta \rho_{r\bm k}(\eta_{in})$ are non-zero initially, where $\eta_{in}$ is an initial moment in conformal time.

Then, the solutions of the perturbation theory equations take the form:

\bea
 \delta \rho_{r\bm k}(\eta)= e^{4 \mathcal H
(\eta_{in}-\eta )} \biggl(4 \sqrt{3} \mathcal H \sin \left(\frac{k
(\eta -\eta_{in})}{\sqrt{3}}\right)\label{22a0}\\+k \cos
\left(\frac{k (\eta
-\eta_{in})}{\sqrt{3}}\right)\biggr)\delta \rho_{r\bm k}(\eta_{in})/k,\\
V_{r\bm k}(\eta)=0,~~~~~~ A_{\bm k}(\eta)=-\frac{B^4 e^{4 \eta
\mathcal H}}{4 \rho_{r0}}\delta \rho_{r\bm k}(\eta),
\label{22a}
\\
F_{\bm k}(\eta)=-\frac{3 B^4 e^{4 \mathcal H \eta_{in}}}{2 k^2
\rho_{r0} \left(3 \mathcal H^2+k^2\right)}\biggl(\left(12 \mathcal
H^2+k^2\right) \cos \left(\frac{k (\eta -\eta_{in}
)}{\sqrt{3}}\right)\nonumber\\+3 \sqrt{3} \mathcal H k \sin
\left(\frac{k (\eta -\eta_{in})}{\sqrt{3}}\right)\biggr)\delta
\rho_{r\bm k}(\eta_{in}),
\label{23a}
\\
V_{v\bm k}(\eta)=\frac{B^2\mathcal H^2{M_p}^2 e^{4\mathcal H
\eta_{in} -2 \eta\mathcal H}}{12 k \rho_{r0} \left(3 \mathcal
H^2+k^2\right)}\biggl(\sqrt{3} \left(12 \mathcal H^2+k^2\right)
\sin \left(\frac{k (\eta
-\eta_{in})}{\sqrt{3}}\right)\nonumber\\-9 \mathcal H k \cos
\left(\frac{k (\eta -\eta_{in})}{\sqrt{3}}\right)\biggr)\delta
\rho_{r\bm k}(\eta_{in}), ~~~~ \delta \rho_{v\bm
k}(\eta)=3{\mathcal H}V_{v\bm k}(\eta).
\label{24a}
\eea

\noindent The quantities $V_{v\bm k}(\eta)$ and $\delta \rho_{v\bm
k}(\eta)$ will not be needed for the CMB spectrum calculations and
will not be considered further.

\subsection{``Gauge Invariant'' Variables}
\label{subsec:15}

The issue is that the metric (\ref{int1}) has not a typical form of
\be
ds^2=a^2(\eta)\left((1+2\Phi(\eta,\bm
x))d\eta^2-\left(1-2\Psi(\eta,\bm
x)\right)\delta_{ij}dx^idx^j\right),
\label{muhmet}
\ee
\noindent which appears in the conventional perturbation theory \cite{dod,mukh} of GR. The comparability of previous results with those of the GR conventional perturbation theory can be provided by the ``gauge invariant'' densities, velocities and potentials \cite{mukh}:
\bea
\tilde \delta_{r\bm k}(\eta )= \frac{\delta \rho_{r\bm k}(\eta
)}{\rho_r(\eta )}-2 \alpha^\prime(\eta ) F_{\bm k}^\prime(\eta
),~~~~~ \tilde v_{r\bm k}=\frac{V_{r\bm k}(\eta )}{\rho_r(\eta
)+p_r(\eta
)}-\frac{F^\prime_{\bm k}(\eta )}{2},\nonumber\\
\Phi_{\bm k}(\eta)=A_{\bm k}(\eta)+\frac{a'(\eta ) F_{\bm k}'(\eta
)+a(\eta ) F_{\bm k}''(\eta )}{2 a(\eta )}, \nonumber\\
\Psi_{\bm k}(\eta)=-\frac{a'(\eta ) F_{\bm k}'(\eta )}{2 a(\eta
)}-A_{\bm k}(\eta )+\frac{1}{6} k^2 F_{\bm k}(\eta).
\label{invq}
\eea
We could not work with the ``invariant'' potentials initially
because the metric (\ref{muhmet}) has not the form (\ref{int}) and
does not admit obtaining the consistent system of the equations
when the zero-order Friedmann equation is violated, i.e.,
satisfied up to some constant (\ref{Fried}). For our simplified
approach, when only initial value of $\delta \rho_{r\bm k}$ is
nonzero, the calculated ``invariant quantities'' are
\bea
\tilde \delta_{r\bm k}(\eta )=\frac{1}{ \left(3 \mathcal
H^2+k^2\right)}\biggl(\left(12 \mathcal H^2+k^2\right) \cos
\left(\frac{k (\eta -\eta_{in})}{\sqrt{3}}\right)\nonumber\\+3
\sqrt{3} \mathcal H k \sin \left(\frac{k (\eta
-\eta_{in})}{\sqrt{3}}\right)\biggr)\delta_{r\bm k}(\eta_{in}),
\nonumber\\
 \tilde v_{r\bm k}(\eta )=\frac{1}{4
k\left(3 \mathcal H^2+k^2\right)}\biggl(9 \mathcal H k \cos
\left(\frac{k (\eta
-\eta_{in})}{\sqrt{3}}\right)\nonumber\\-\sqrt{3} \left(12
\mathcal H^2+k^2\right) \sin \left(\frac{k (\eta
-\eta_{in})}{\sqrt{3}}\right)\biggr)\delta_{r\bm k}(\eta_{in}),\nonumber\\
\Phi_{\bm k}(\eta)=0,~~~~~\Psi_{\bm k}(\eta)=0,
\label{inv}
\eea
where we take into account that $\frac{\rho_{r0}}{B^4\exp\left(4
\mathcal H \eta_{in} \right)}=\rho_r(\eta_{in})$ and $\delta_{r\bm
k}(\eta_{in})=\frac{\delta \rho_{r\bm
k}(\eta_{in})}{\rho_r(\eta_{in})}$.
The potentials $\Phi_{\bm k}$, $\Psi_{\bm k}$ are zero only because
we use the simplified initial condition, where $\delta \rho_{r\bm
k}$ is nonzero initially.

\subsection{Silk Dumping}
Electrons scatter the photons before the time of the last
scattering surface. Although we consider photon-electron-baryon
plasma as some perfect medium with the equation of state $w=1/3$,
the photon diffusion due to the Thompson scattering contributes to
the electron-photon scattering process \cite{CMB}. To estimate
this (so-called \emph{Silk dumping}) contribution to the
perturbations,
 we follow the methodology of Refs. \cite{dod,mukh} suggesting the suppression
 of the expressions (\ref{22a0}),(\ref{22a}),(\ref{23a}),(\ref{24a}) and (\ref{invq}) by the factor
$\exp\left(-k^2/k_D^2\right)$, where $k_D$ is written as
\cite{mukh}
\be
k_D(\eta_r)\approx\left(\frac{2}{15}\int_0^{\eta_r}\frac{d\eta}{\sigma_T
n_e a}\right)^{-1/2}=\left(\frac{2}{15\sigma_T
n_{b0}}\int_0^{\eta_r}a^2d\eta\right)^{-1/2},
\label{kD}
\ee
and $\sigma_T=6.65\times10^{-25}~cm^2$ is the Thompson cross
section. The free electron density $n_e$ before recombination equals
to the baryon density and scales as $n_e=n_{b0} a^{-3}$, where
$n_{b0}$ is the baryon present density
\be
n_{b0}=\Omega_b \frac{M_p^2\mathcal H^2}{2m_p}
\label{bar}
\ee
expressed through a dimensionless quantity $\Omega_b$, a proton
mass $m_p$ and a critical density $M_p^2\mathcal H^2/2$. Formally,
for the dependence given by (\ref{aeta}), an integration in
(\ref{kD}) has to begin from $\eta=-\infty$. However, as was shown
in \cite{our1}, the universe started from a power-law expansion
changed by (\ref{aeta}) afterward. It was also shown, that $B$ is
of the order of $10^{-30}$. Under this condition, $B$ does not
play a role if the lower
 limits of $\eta$ equal $-\infty$ or zero (the results are approximately the same in both cases).

Substituting the dependence (\ref{aeta})
 and the conformal time of the last scattering surface
$\eta_r=\frac{1}{\mathcal H}\ln \frac{10^{-3}}{B} $, that corresponds
to the scale factor $a_r\approx 10^{-3}$, into (\ref{kD}) results in
\be
k_D(\eta_r)=\sqrt{15\sigma_T n_{b0}\mathcal H}\times10^3\approx
10^3\sqrt{\Omega_b}\, \mathcal H.
\label{silk}
\ee
As one may see, plasma is closer to an ideal fluid for greater matter density. For instance, the conventional value of $\Omega_b=0.03$ results in the damping scale of $k_D\sim 170$ in the units of
$\mathcal H$.

\section{CMB Spectrum}

In the previous section, we have considered the perturbation theory which describes the evolution of the plasma (radiation) in the presence of the vacuum perturbations. This evolution extends up to the ``last scattering surface'', i.e., up to a moment when the universe becomes transparent for radiation. Conformal time of the last scattering surface $\eta_r$ corresponds to the temperatures $T_r\sim 3000 ~K $ and the redshift $z_r\approx 1100$.  Describing the photons' propagation from the last scattering surface to an observer is insufficient to use hydrodynamic approximation so that the Boltzmann equation is needed, which can be written in the form of
\be \frac{\ptl f}{\ptl \eta}+\frac{d x^i}{d\eta}\frac{\ptl f}{\ptl
x^i}+\frac{d p_i}{d\eta}\frac{\ptl f}{\ptl p_i}=St[f],
\ee
where the right hand side  $St[f]$ represents the collision
integral.
 If the distribution function $f$ is assumed to be a scalar, it would depend  on $x^i$
 and $p_i$ because
the photon number $dN=f(x^i,p_j,\eta)dx^1dx^2dx^3dp_1dp_2dp_3$ is
scalar according to the Liouville theorem and the quantity
$dx^1dx^2dx^3dp_1dp_2dp_3$ is scalar. The expressions describing
the photon propagation are
\bea
\frac{d p^\alpha}{d \lambda}=-\Gamma_{\beta\gamma}^\alpha p^\beta
p^\gamma=-\Gamma_{\beta\gamma}^\alpha
g^{\beta\sigma}g^{\gamma\delta} p_\sigma p_\delta,\nonumber\\
\frac{d x^\alpha}{d \lambda}=p^\alpha=g^{\alpha\beta}p_\beta,
\eea
where $\lambda$ is an affine parameter along the photon
trajectory. Using the last equation for the zero component  $\frac{d
x^0}{d \lambda}=\frac{d \eta}{d \lambda}=p^0$ of derivatives with respect to
$\lambda$ allows rewriting it in the terms of derivatives with respect to $\eta$.

Then, the Boltzmann equation can be reduced to the equation for a temperature perturbation by substitution
\be
 f(x^i,p_j,\eta)=\frac{1}{\exp\left(\frac{p_0(\eta)}{T_0(\eta)\sqrt{g_{00}}\left(1+\Theta(\bm n,\bm x,
 \eta)\right)}\right)-1},
\ee
where $\Theta(\bm n,\bm x, \eta)$ is a temperature contrast and a unit vector $n^i=p_i/(\sum_{n=1}^3 p_n^2 )$.
 Finally, for the coefficients of the Fourier transform $\Theta(\bm n,\bm x,
 \eta)=\sum_{\bm k}\Theta_{\bm k}(\eta,\bm n)e^{i\bm k \bm x}$ calculations with the metric (\ref{int1})
give
\be
\frac{\ptl \Theta_{\bm k}}{\ptl \eta}-i k \mu \Theta_{\bm k}-i k
\mu A_{\bm k}+A_{\bm
k}^\prime+\frac{k^2}{6}\left(3\mu^2-1\right)F_{\bm
k}^\prime=\tau^\prime(\Theta_{\bm k}-\Theta_{0\bm k}- v_{b\bm
k}\,\mu),
\label{col}
\ee
where $\mu=\bm n\cdot\bm k/k$ is the cosine of the angle between
$\bm n$ and $\bm k$, $\Theta_{0\bm k}(\eta)$ is the component
$l=0$ of $\Theta_{\bm k}(\bm n,\eta)$  in the expansion of the
Legendre polynomials
\be
\Theta_{l\bm k}=i^l\int_{-1}^1P_l(\mu)\Theta_{\bm
k}(\mu)\frac{d\mu}{2},
\label{expl}
\ee
and $v_{b\bm k}$ is the Fourier transform of the function
determining baryon velocity. The function $\tau(\eta)$ describes
the photon Compton scattering by electrons: $\tau^\prime=-\sigma_T
n_e a$, where $\sigma_T$ is a cross section of the Thomson
scattering and $n_e$ is a free electron density. Before the last
scattering surface, the photons are tightly coupled with electrons
and protons by the Thomson scattering, and the electrons, in turn,
are tightly coupled with baryons by the Coulomb interaction. As a
consequence, any bulk motion of the photons must be shared by the
baryons. Although we do not consider baryons explicitly, one may
assume roughly that
 baryons and photons are in equilibrium and thus \cite{dod}
\be
v_{b\bm k}=-3 i \Theta_{1\bm k}(\eta).
\label{32}
\ee
Further, the monopole $\Theta_{0\bm k}$ and dipole $\Theta_{1\bm k}$ components of the
temperature perturbations can be connected with the perturbations of density and velocity. From one hand side, the
{00}-component of the energy-momentum tensor in line with (\ref{tmn}) is
\be
T_{0\bm k}^0=\delta \rho_{\bm k}(\eta).
\label{33}
\ee
On the other hand, it can be expressed via a temperature perturbation \cite{mukh}:
\be
T^0_{0\bm k}=4\rho_r\int \Theta_{\bm k}(\bm n,\eta)\frac{d^2\bm
n}{4\pi}.
\label{34}
\ee

Comparison of (\ref{33}) and (\ref{34}) gives $\Theta_{0\bm
k}(\eta)=\frac{1}{4\rho_r}\delta \rho_{r\bm
k}(\eta)=\frac{1}{4}\delta_{r\bm k}.$ Analogously, in the first order of
the perturbation theory, the components $T_{0i}$ take the form
\be
T_{0j}=-a^2(\eta)(\rho_r(\eta)+p_r(\eta))\ptl_j v_r(\eta,\bm x),
\ee
or
\be
T_{0\bm k}^{j}=\frac{4}{3}\rho_r(\eta)i k_j v_{r\bm k}(\eta).
\label{37}
\ee
\noindent At the same time \cite{mukh}
\be
T_{0\bm k}^j=-4\rho_r\int n^j \Theta_{\bm k}(\bm n,\eta)
\frac{d^2n}{4\pi}.
\label{38}
\ee

As consequence of (\ref{expl}), (\ref{32}), (\ref{37}) and (\ref{38}), one has $v_{b\bm k}=-3i \Theta_{1\bm k}=-ikv_{r\bm k}$,
and Eq. (\ref{col}) can be rewritten in the form of
\be
\Theta_{\bm k}^\prime-(i k \mu+\tau^\prime)\Theta_{\bm k}=e^{i
k\mu \eta+\tau}\frac{d}{d\eta}\left(\Theta_{\bm k} e^{-i k\mu
\eta-\tau} \right)=S_{\bm k},
\label{89}
\ee
where $S_{\bm k}=-\tau^\prime\frac{\delta_{r\bm k}}{4}+\tau^\prime
i k \mu\,v_{r\bm k}+ik \mu A_{\bm k} -A_{\bm k}^\prime
-\frac{k^2}{6}\left(3\mu^2-1\right)F_{\bm k}^\prime.$

Solution of Eq. (\ref{89}) takes the form of
\bea
\Theta_{\bm k}(\eta_0)=\Theta_{\bm k}(\eta_{in})e^{-i\mu
k(\eta_{in}-\eta_0)-\tau(\eta_{in})+\tau(\eta_0)}+
\int_{\eta_{in}}^{\eta_0} S_{\bm k} \,e^{-i\mu k(\eta-\eta_0)-\tau
(\eta)+\tau (\eta_0)}d\eta\approx\nonumber
\\\int_{\eta_{in}}^{\eta_0}
e^{-\tau(\eta)}\biggl(-\tau^\prime\frac{\delta_{r\bm
k}}{4}-\tau^\prime v_{r\bm k}\frac{d}{d\eta}-A_{\bm
k}^\prime-A_{\bm
k}\frac{d}{d\eta}~~~~~~~~~~~~~~~\nonumber\\-\frac{ F_{\bm
k}^\prime}{6}\left(-3\frac{d^2}{d\eta^2}-k^2\right)\biggr)e^{-i
k\mu (\eta-\eta_0)}d\eta,~~~~~~
\eea
where $\eta_0$ is the present day conformal time, $\eta_{in}$ is
some initial moment of time before the last scattering surface,
when the universe was not transparent for light. The terms
containing $e^{-\tau(\eta_{in})}$ are omitted because the
function $e^{-\tau(\eta)}$ vanishes quickly if $\eta<\eta_r$
\cite{dod}.

Using the integral (\ref{expl}) and the integral
\be
\int_{-1}^1\frac{d\mu}{2}P_l(\mu)e^{-ik\mu(\eta-\eta_0)}=\frac{1}{i^l}j_l(k(\eta-\eta_0))
\ee
leads to
\bea
\Theta_{l\bm k}(\eta_0)=\int_{\eta_{in}}^{\eta_0}
e^{-\tau(\eta)}\biggl(\left(-\tau^\prime \frac{1}{4}\delta_{r\bm
k}-A_{\bm k}^\prime+\frac{F_{\bm k}^\prime
k^2}{6}\right)j_l(k(\eta-\eta_0)) ~~~~~~\nonumber\\
-\left(\tau^\prime v_{r\bm k}+A_{\bm k}\right)k
j_l^\prime(k(\eta-\eta_0))+\frac{F_{\bm
k}^\prime}{2}k^2j_l^{\prime\prime}(k(\eta-\eta_0))
 \biggr)d\eta.
 \label{finthet}
\eea

One may rewrite Eq. (\ref{finthet}) in the terms of invariant potentials, densities and velocities (\ref{invq}):

\bea
\Theta_{l\bm
k}(\eta_0)=\int_{\eta_{in}}^{\eta_0}e^{-\tau(\eta)}\biggl(-\tau^\prime
\bigl(\frac{ \tilde \delta_{r\bm k}}{4}+\Phi_{\bm k}
\bigr)j_l(k(\eta-\eta_0))~~~~~~~~\nonumber\\-\tau^\prime\tilde
v_{r\bm k}k\,j^\prime_l(k(\eta-\eta_0))+(\Phi_{\bm
k}^\prime+\Psi_{\bm k}^\prime)j_l(k(\eta-\eta_0))\biggr)d\eta.
\label{finthet2}
\eea
The integrand expressions  in (\ref{finthet}) and
(\ref{finthet2}) differ by a total derivative, which does not contribute to the integral because $e^{-\tau(\eta_{in})}\approx 0$ at the lower limit, and the Bessel function $j_l(0)=0$
for $l>0$ at the upper limit.

According to (\ref{inv}), the invariant potentials $\Psi$ and $\Phi$ equal zero in our simplified consideration when only $\delta_{r\bm k}(\eta_{in})$ is nonzero. Thus, there is no the Sachs-Wolf effect \cite{CMB} and the expression (\ref{finthet2}) is reducible to

\bea
\Theta_{l\bm
k}(\eta_0)=\int_0^{\eta_0}(-\tau^\prime)e^{-\tau(\eta)}\biggl(
\frac{ \tilde \delta_{r\bm k}}{4} j_l(k(\eta-\eta_0))+\tilde
v_{r\bm
k}k\,j^\prime_l(k(\eta-\eta_0))\biggr)d\eta\nonumber\\\approx
\frac{ \tilde \delta_{r\bm k}(\eta_r)}{4}
j_l(k(\eta_r-\eta_0))+\tilde v_{r\bm
k}(\eta_r)k\,j^\prime_l(k(\eta_r-\eta_0)),
\label{finthet2}
\eea
where the fact is used that the visibility function
$g(\eta)=-\tau^\prime e^{-\tau(\eta)}$ \footnote{The visibility function gives the probability of a CMB photon scattering out of the line of sight
within of a $d\eta-$layer on the last scattering surface \cite{mukh}.} is peaked near last
scattering surface $\eta_r$. On the other hand, the integral $\int g(\eta)d\eta=1$, and thereby, it is like the Dirac delta-function
$g(\eta)=\delta(\eta-\eta_r)$.

Using the expressions for $\tilde \delta_{r\bm k}$ and $\tilde
v_{r\bm k}$ from (\ref{inv}), we obtain the expressions for the coefficients
\bea
C_l=\frac{2}{\pi}\int_0^\infty   <\Theta_{l\bm k}(\eta_0)>^2k^2
dk~~~~~~~~~~~~~~~~~~~~~~~~~~
~~~~~~~~~~~~~~~~~~~~~~~~\nonumber\\
=\frac{2}{\pi}\int_0^\infty \Biggl| \frac{\left(12\mathcal
H^2+k^2\right) \cos \left(\frac{k (\eta_r
-\eta_{in})}{\sqrt{3}}\right)+3 \sqrt{3}\mathcal H k \sin
\left(\frac{k (\eta_r -\eta_{in})}{\sqrt{3}}\right)}{4 \left(3
\mathcal H^2+k^2\right)}j_l(k(\eta_r-\eta_0))\nonumber\\
+\frac{9\mathcal H k \cos \left(\frac{k (\eta_r
-\eta_{in})}{\sqrt{3}}\right)-\sqrt{3} \left(12\mathcal
H^2+k^2\right) \sin \left(\frac{k (\eta_r
-\eta_{in})}{\sqrt{3}}\right)}{4 \left(3
\mathcal H^2+k^2\right)}j_l^\prime(k(\eta_r-\eta_0))\Biggl|^2\nonumber\\
\mathcal P(k,\eta_{in}) \frac{dk}{k},~~~~\label{50}
\eea
where $\mathcal P(k,\eta_{in})=k^3<\delta_{r\bm
k}(\eta_{in})\delta^*_{r\bm k}(\eta_{in})> $ is a primordial fluid
spectrum which serves as an initial condition for the plasma
perturbations considered in the previous section.

\subsection{Effect of the Finite Thickness of the Last Scattering Surface}
\label{st}
A real-world visibility function $g(\eta)$ is not exactly the Dirac delta-function, but
it is smeared over a finite region of $\eta$.
One may approximately assume that it has the Gaussian form

\be
g(\eta)=-\tau^\prime(\eta)\exp(-\tau)=\frac{1}{\Delta
\eta_r\sqrt{2\pi}}\exp\left(-\frac{(\eta-\eta_r)^2}{2\Delta
\eta_r^2}\right),
\label{gau}
\ee
where $\Delta\eta_r$ is a width of the last scattering surface.
That corresponds to
\be
\tau(\eta)=-\ln\left(\frac{1}{2}+\frac{1}{2}\mbox{erf}\left(\frac{\eta-\eta_r}{\sqrt
2\,\Delta\eta_r}\right)\right).
\ee
Let us consider the exact integral
\be
\int_{-\infty}^\infty g(\eta)e^{i
k(\eta-\eta^*)}d\eta=\exp\left(-k^2\Delta \eta_r^2/2\right)e^{i
k(\eta_r-\eta^*)}. \label{53}
\ee

\begin{figure}[th]
  \includegraphics[width=12cm]{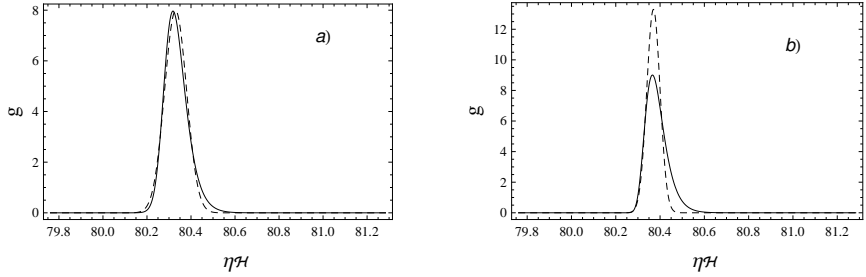}\\
  \caption{
   Visibility function $g(\eta)=-\tau^\prime(\eta)\exp(-\tau)$ for different baryon density (\ref{bar}) a)
   $\Omega_b=0.03$,
   b)
   $\Omega_b=0.3$ (solid curves). Dashed curves are Gaussian approximations (\ref{gau}) with a) $\Delta \eta_r=0.05$ and b) $\Delta \eta_r=0.03$.  }\label{fig1}.
\end{figure}

\begin{figure}[th]
  \includegraphics[width=7cm]{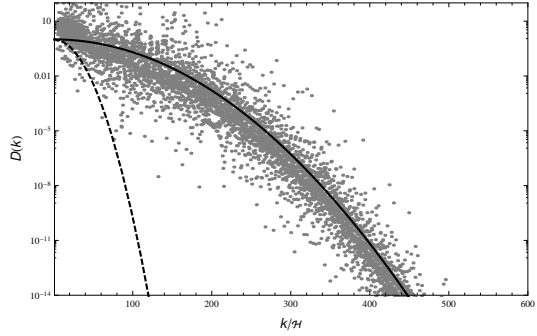}\\
  \caption{Calculated damping factor due to finite width of the last
  scattering surface
   $D(k)=\left(\int_0^{\eta_0}g(\eta)\biggl( \frac{ \tilde
\delta_{r\bm k}(\eta)}{4} j_l(k(\eta-\eta_0))+\tilde v_{r\bm
k}(\eta)k\,j^\prime_l(k(\eta-\eta_0))\biggr)d\eta\right)^2/
\biggl( \frac{ \tilde \delta_{r\bm k}(\eta_r)}{4}
j_l(k(\eta_r-\eta_0))+\tilde v_{r\bm
k}(\eta_r)k\,j^\prime_l(k(\eta_r-\eta_0))\biggr)^2$ for $\Delta
\eta_r=0.03$, $l=300$. Dashed and solid curves correspond to
$D(k)=\exp\left(-(k+k/\sqrt{3})^2\Delta \eta_r^2\right)$ and
$D(k)=\exp\left(-(k-k/\sqrt{3})^2\Delta \eta_r^2\right)$
respectively.
   }\label{fig1a}.
\end{figure}

As it is seen (Eq. \ref{53}), the variable $\eta$ is changed by $\eta_r$ in the expression $e^{i
k(\eta-\eta^*)}$ after integration, and besides a suppression factor appears.

The expression (\ref{finthet2}) contains the exponents $e^{i(k\pm
k/\sqrt{3})\eta}$ originating from both Bessel functions and
$\tilde \delta_{\bm k}$. Thus, the suppression factor $e^{-(k\pm
k/\sqrt{3})^2\Delta \eta_r^2/2}$ appears in (\ref{finthet2}) as a
result of integration, which has to be introduced into the
integrand of (\ref{50}). The overall damping factor originates
from both Silk dumping and finite width of the last scattering
surface, but the last gives the main contribution. The calculation
of the last scattering surface width has to take into account the
process of hydrogen recombination. In the standard $\Lambda$CDM
model, one needs using the
 kinetic equations involving at
least three levels of the hydrogen atom. The Milne-like universe
expands at $\sqrt{z_r}-$ times slower than the standard
$\Lambda$CDM one. Thus, the Saha equilibrium equation
\cite{mukh,dod}
\be
\frac{n_p n_e}{n_H}=\frac{X_e^2}{1-X_e}n_{b}=\left(\frac{T
m_e}{2\pi}\right)^{3/2}\exp\left(-\frac{B_H}{T}\right)
\label{saha}
\ee
is a good estimation, where $n_p$ is a proton density, and $n_H$ is
a density of neutral atoms.

Eq. (\ref{saha}) allows obtaining the hydrogen ionization degree $X_e=n_p/n_{b}$, where $n_{b}=n_p+n_H$.
An optical depth \cite{mukh} is calculated as
\be
\tau(\eta)=\sigma_T\int_\eta^{\eta_0}n_{b}(\eta^\prime)X_e(\eta^\prime)a(\eta^\prime)d\eta^\prime,
\ee
where $n_b$ scales as $n_b(\eta)=n_{b0}/a^3(\eta)$ and $n_{b0}$ is
given by (\ref{bar}).  The visibility function for different
values of the matter density is shown in Fig. \ref{fig1}. As one
can see the  width $\Delta \eta_r$ of the Gaussian approximation
is $0.05$ for  $\Omega_b=0.03$ and $0.03$ for $\Omega_b=0.3$.
In the last case, the visibility function has
non-Gaussian shape. However, the initial stage of recombination
affects mainly the ``left front'' of the visibility function which
becomes ``sharper'' and can be approximated by a Gaussian function
shown in Fig. \ref{fig1} b.

The expression (\ref{53}) is exact only for averaging of the
exponent, however it is approximately valid and for more
complicated expressions like  the integrand of (\ref{finthet2}). As one can see from Fig. \ref{fig1a}, the lowest
suppression factor  $e^{-(k- k/\sqrt{3})^2\Delta \eta_r^2/2}$
should be taken for the calculations.

\section{Results and discussion}

A distance from the last scattering surface to the present time
observer is $\eta_0-\eta_r$. For the Milne-like universe
(\ref{aeta}) these distances are $\eta_0=\frac{1}{\mathcal
H}\ln\frac{1}{B}$ and $\eta_r=\frac{1}{\mathcal
H}\ln\frac{a_r}{B}$ respectively. Thus, one has $\eta_0-\eta_r\sim
\mathcal H^{-1}\ln z_r\sim 7\mathcal H^{-1}$ independent of $B$.

\begin{figure}[th]
  \includegraphics[width=7cm]{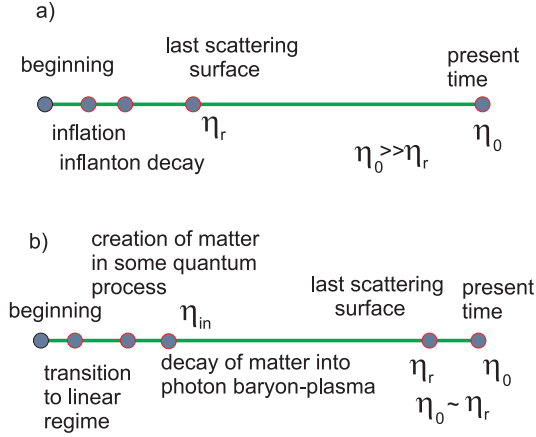}\\
  \caption{
   Schematic representation of the time scales in
the a) standard $\Lambda$CDM and b) linear cosmologies
respectively.}
  \label{fig2}
\end{figure}

To calculate the spectrum according to (\ref{50}), one needs
knowing the initial spectrum.  The standard model of cosmological
inflation gives almost flat spectrum, i.e., $\mathcal P(k)\approx
const$ and the oscillations in the observed CMB anisotropy
spectrum are interpreted as a result of acoustic oscillation of
the photon-baryon plasma. There is a principled difference between
the standard model and the linear cosmology considered here. In
the standard model, the typical angular scale is $\theta\sim
\frac{\eta_r-\eta_{in}}{\eta_0-\eta_{r}}\sim
\frac{\eta_r}{\eta_0}$. As a consequence of $\eta_r<<\eta_0$ in
the $\Lambda$CDM model, one may obtain the angular scale of
$\theta\sim 1^o$ coinciding with the experimental one. In the
linear cosmology $\eta_r\sim \eta_0$ (see scheme in Fig.
\ref{fig2}) and the spectrum oscillations should have another
origin. In particular, they could originate from the oscillations
of the initial spectrum $\mathcal P(k,\eta_{in})$, which can
 be taken in
the form
\be
\mathcal P(k,\eta_{in})=3\times 10^{-7}|\sin k\eta_{in}|^2.
\label{rr}
\ee
For the dependence (\ref{aeta}), one has to take $\eta_{in}\sim 0.06$ to obtain experimentally observed
angular scale, that gives
$\theta \sim \frac{\eta_{in}}{\eta_0-\eta_r}\sim 0.4^o$.

 It is easy to calculate cosmic (i.e. physical) time $t_{in}$ corresponding to the
conformal time $\eta_{in}$. Integrating with (\ref{aeta}) gives
$t_{in}=\int_0^{\eta_{in}}a(\eta)d \eta\approx B\,\eta_{in}$,
where it is taken into account that $\mathcal H\eta_{in}\ll 1$.
For instance, taking $\eta_{in}=0.06/\mathcal H$ and
$B=3.8\times10^{-38}$ gives $t_{in}=5.9\times 10^{-22}~s$, which
corresponds to the lifetime of the Higgs boson
$t_H=2\pi/\Gamma_H$, where $\Gamma_H=7~MeV$. Here, it is implied
that Higgs bosons are created initially \cite{our1}, then decay
into another particles and, finally, into the baryons and photons.
Taking another value of $B$ one requires connecting $\eta_{in}$
with another physical process.

\begin{figure}[th]
  \includegraphics[width=11cm]{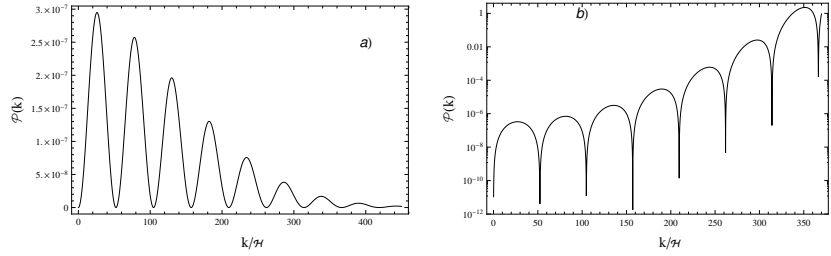}\\
  \caption{
  a) Initial spectrum multiplied by the all the damping factors, i.e., the resulting
  spectrum $\mathcal P(k)=3\times 10^{-7}|\sin 0.06 k|^2\exp(-k^2/200^2)$, which reproduces the observational data qualitatively.  b) Rising initial spectrum
$\mathcal P(k)=3\times 10^{-7}|\sin 0.06 k|^2\exp(k^2/87^2)$. It
is seen, that the perturbations
   with $k>350\mathcal H$ lie in the nonlinear region, because $\mathcal P(k)>1$.}
  \label{fig2}
\end{figure}

\begin{figure}[th]
  \includegraphics[width=11cm]{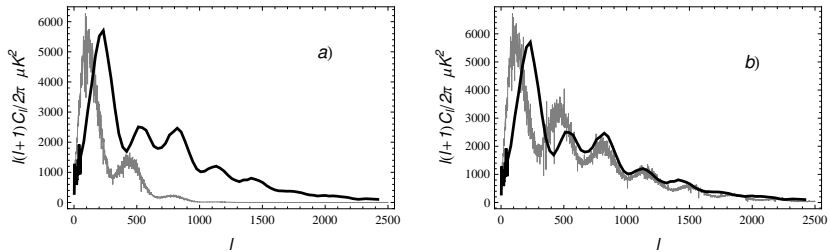}\\
  \caption{
   Cosmic microwave anisotropy spectrum calculated within the framework of
the linear Milne-like cosmology (gray noisy curve).  Black curve corresponds to the
   ``Planck''-satellite  data \cite{planck}. The
quantities $C_l$ are dimensionless (the multiplication by the squared
present CMB temperature gives the dimensional $C_l$). a) corresponds to
the initial spectrum (\ref{rr}), b) corresponds to the rising
spectrum (\ref{rrs}). }
  \label{fig3}
\end{figure}

The initial spectrum (\ref{rr}) has to be multiplied by the
damping factor\footnote{The case of the best agreement with the observational data is considered:
$\Omega_m=0.3$, $\Delta \eta_r=0.03$.}
\be
D(k)\approx\exp\left(-(k-k/\sqrt{3})^2\Delta
\eta_r^2\right)\approx\exp\left(-k^2/80\right).
\ee and substituted into Eq.
(\ref{50}). We do not predict absolute values, and the coefficient
in (\ref{rr}) is taken to reproduce only highest first CMB
peak. The result, shown in Fig. \ref{fig3} (a) demonstrates a too strong
suppression of higher harmonics in comparison with the observational data. To improve the agreement, one may take a rising initial spectrum
\be
\mathcal P(k,\eta_{in})=3\times 10^{-7}|\sin
k\eta_{in}|^2\exp\left(k^2/\kappa_{in}^2\right).
\label{rrs}
\ee
with $\kappa_{in}=87 $ in order to obtain the overall damping factor
 about of
$\exp\left(-k^2/200^2\right)$, because $80^{-2}-87^{-2}\approx
200^{-2}$.

 The result of
calculation with this formula is shown in Fig. \ref{fig3} (b). The
\emph{Planck}-satellite data give a very precise measurement of
the CMB anisotropy \cite{CMB,esa,planck,hu}. One can see the
qualitative coincidence with the spectrum observed by the
\emph{Planck}-satellite. The positions of the peaks are shifted
relatively observed ones. However, it is no wonder because the
model considered is rough and requires further development. At
least, the model needs taking into account the baryonic content
explicitly. Of course, no analytic solutions for perturbations
could be found with this complication. The Silk dumping and the
finite width of the last scattering surface have to be taken into
account more accurately. Besides, more complicated models of the
initial spectrum have to be considered.

From a fundamental point of view, it could imagine
 some breathtaking physics like the inflation theory. However, it could be quite different,
  because the inflation cannot produce, a ``violet'', i.e., rising with $k$, initial spectrum (\ref{rrs}).
  In principle, the linear cosmology needs no inflation, because the scales of perturbations modes
  always remain within the horizon and there is no need in any model like inflation for the superhorizon
  spectral modes. Thus, the liner universe seems in some sense  simpler compared to the standard $\Lambda$CDM model. However,
  the most fundamental problem of the linear cosmology is a requirement of more accurate consideration of vacuum
  perturbations with taking into account the quantum properties of the vacuum. The above simple model of vacuum as a
  fluid with the equation of state $w=-1/3$ is an only very rough heuristic approximation.

Unfortunately, well-known software packages such as CAMB \cite{camb} and
CMBFAST \cite{cmbfast} are absolutely useless for the calculation
of CMB spectrum in the linear cosmology because they assume a
quite different formation mechanism for the CMB spectrum peaks. It
seems that the tools for the ionization history analysis, such as
RECFAST \cite{rec}, also have to be modified to take into account
more than three levels of the hydrogen atom. It results from the
fact that partially ionized hydrogen plasma is closer to thermal
equilibrium due to the slower expansion of the Milne-like universe
and, thereby, more hydrogen levels are populated. It seems that
the pure equilibrium Saha formula used above gives a sufficiently
good approximation in this case.

It should also to do some notes about distortion of the CMB
spectrum from blackbody one \cite{dist}.  The expected distortion
of the spectrum caused
 by hydrogen recombination should be mach smaller than that  in the
$\Lambda$CDM model.

\section*{Appendix}
\addcontentsline{toc}{section}{Appendix}

The expression for the perturbation of the vacuum density is given
by
\bea
\delta \rho_{v\bm k}=\frac{B^2 \mathcal H^3 {M_p}^2 e^{-4 \eta
\mathcal H}}{8 {\rho_{r0}}^2 \left(3 \mathcal
H^2+k^2\right)}\biggl(-C_1\bigl(B^2 \mathcal H {M_p}^2 e^{2 \eta
\mathcal H} (3 \mathcal H^2+2 i \sqrt{3} \mathcal H
k-k^2)\nonumber\\-6 \mathcal H {\rho_{r0}}+2 i \sqrt{3} k
{\rho_{r0}}\bigr)e^{-i \frac{ \eta k}{\sqrt{3}}}+C_2\bigr(-B^2
\mathcal H {M_p}^2 e^{2 \eta  \mathcal H} (3 \mathcal H^2-2 i
\sqrt{3} \mathcal H k-k^2)\nonumber\\+6 \mathcal H {\rho_{r0}}+2 i
\sqrt{3} k {\rho_{r0}}\bigl)e^{i \frac{ \eta
k}{\sqrt{3}}}\biggr)+\frac{B^2 \mathcal H^3 {M_p}^2 e^{-2 \eta
\mathcal H}}{4{\rho_{r0}} \left(3 \mathcal
H^2+k^2\right)}\biggl(C_3(3 \mathcal H+i \sqrt{3} k)e^{-i \frac{
\eta k}{\sqrt{3}}}\nonumber\\+C_4(3 \mathcal H-i \sqrt{3} k)e^{i
\frac{ \eta k}{\sqrt{3}}}\biggr)-\frac{k^4 {M_p}^2 e^{-3 \eta
\mathcal H}}{18 B^2}\biggl(Z_1e^{-\eta \sqrt{\mathcal
H^2+\frac{k^2}{3}}}+Z_2e^{\eta \sqrt{\mathcal
H^2+\frac{k^2}{3}}}\biggr),\nonumber
\eea

Then
\bea
 V_{v\bm k}= -\frac{B^2 \mathcal H^2 M_p^2 e^{-4 \eta \mathcal H}}{24\rho_{r0}^2
\left(3 \mathcal H^2 k+k^3\right)}\biggl(C_1\bigl(B^2 \mathcal H k
{M_p}^2 e^{2 \eta \mathcal H} (3 \mathcal H^2+2 i \sqrt{3}
\mathcal H k-k^2)\nonumber\\-2 i {\rho_{r0}} (6 \sqrt{3} \mathcal
H^2-3 i \mathcal H k+\sqrt{3} k^2)\bigr)e^{-i \frac{ \eta
k}{\sqrt{3}}}+C_2\bigl(B^2 \mathcal H k {M_p}^2 e^{2 \eta \mathcal
H} (3 \mathcal H^2\nonumber\\-2 i \sqrt{3} \mathcal H k-k^2)+2 i
{\rho_{r0}}(6 \sqrt{3} \mathcal H^2+3 i \mathcal H k+\sqrt{3}
k^2)\bigr)e^{i \frac{ \eta
k}{\sqrt{3}}}\biggr)\nonumber\\
+\frac{B^2 \mathcal H^2{M_p}^2 e^{-2 \eta  \mathcal H}}{12
{\rho_{r0}} \left(3 \mathcal H^2+k^2\right)}\biggl(C_3(3 \mathcal
H+i \sqrt{3} k)e^{-i \frac{ \eta k}{\sqrt{3}}}+C_4(3 \mathcal H-i
\sqrt{3} k)e^{i \frac{ \eta
k}{\sqrt{3}}}\biggr)\nonumber\\+\frac{k^2 {M_p}^2 e^{-3 \eta
\mathcal H}}{54 B^2}\biggl(Z_1\bigl(\sqrt{3} \sqrt{3 \mathcal
H^2+k^2}+3\mathcal H\bigr)e^{-\eta \sqrt{\mathcal
H^2+\frac{k^2}{3}}}\nonumber\\+Z_2\bigl(\sqrt{3} \sqrt{3 \mathcal
H^2+k^2}-3\mathcal H\bigr)e^{\eta \sqrt{\mathcal
H^2+\frac{k^2}{3}}}\biggr),\nonumber
\\
F_{\bm k}= \frac{B^4 e^{-2 \eta  \mathcal H}}{4 {\rho_{r0}}^2
\left(3 \mathcal H^2 k+k^3\right)}\biggl(C_1\bigr(B^2 \mathcal H
{M_p}^2 e^{2 \eta \mathcal H}(-3 i \sqrt{3} \mathcal H^2+6
\mathcal H k\nonumber\\+i \sqrt{3} k^2)-6 i \sqrt{3} \mathcal H
\rho_{r0}-6 k \rho_{r0}\bigl)e^{-i \frac{ \eta
k}{\sqrt{3}}}+C_2\bigl(B^2 \mathcal H {M_p}^2 e^{2 \eta  \mathcal
H} (3 i \sqrt{3} \mathcal H^2\nonumber\\+6 \mathcal H k-i \sqrt{3}
k^2)+6 i \sqrt{3} \mathcal H \rho_{r0}-6 k \rho_{r0}\bigr)e^{i
\frac{ \eta k}{\sqrt{3}}}\biggr) -\frac{3 B^4}{2 k \rho_{r0}
\left(3 \mathcal H^2+k^2\right)}\nonumber\\\biggl(C_3(k-i \sqrt{3}
\mathcal H)e^{-i \frac{ \eta k}{\sqrt{3}}}+C_4(k+i \sqrt{3}
\mathcal H)e^{i \frac{ \eta k}{\sqrt{3}}}\biggr)+e^{-\mathcal
H\eta}\biggl(Z_1 e^{-\frac{\eta \sqrt{3 \mathcal
H^2+k^2}}{\sqrt{3}}}\nonumber\\+Z_2 e^{\frac{\eta \sqrt{3 \mathcal
H^2+k^2}}{\sqrt{3}}}\biggr),\nonumber
\\
A_{\bm k}=\frac{e^{-2 \eta  \mathcal H}}{24
\rho_{r0}^2}\biggl(C_1\bigl(-6 B^4 \rho_{r0}+B^6 \mathcal H
{M_p}^2 e^{2 \eta  \mathcal H} (3 \mathcal H+i \sqrt{3}
k)\bigr)e^{-i \frac{ \eta k}{\sqrt{3}}}\nonumber\\+C_2\bigl(-6 B^4
\rho_{r0}+B^6 \mathcal H{M_p}^2 e^{2 \eta  \mathcal H} \left(3
\mathcal H-i \sqrt{3} k\right)\bigr)e^{i \frac{ \eta
k}{\sqrt{3}}}\biggr)\nonumber\\-\frac{B^4}{4 \rho_{r0}}\biggl(C_3
e^{-\frac{i \eta  k}{\sqrt{3}}}+C_4 e^{\frac{i \eta
k}{\sqrt{3}}}\biggr).\nonumber
\eea

%
%
%

\end{document}